\documentstyle[12pt]{article}
\newcommand{\be}{\begin{equation}}
\newcommand{\ee}{\end{equation}}
\newcommand{\ba}{\begin{eqnarray}}
\newcommand{\ea}{\end{eqnarray}}
\newcommand{\mj}{\mid_{\sigma=\sigma_{0},\vec\pi=0}}
\def\mjs{\mid_{\sigma=\sigma_0(T),\vec\pi=0}}
\newcommand{\m}{\mid_{\sigma=\sigma_{c},\vec\pi =\vec\pi_{c}}}
\newcommand{\no}{\nonumber \\}
\begin{document}

\begin{titlepage}\begin{center}

\hfill{HYUPT-95/04}

\vskip 0.4in
{\Large\bf Finite-Temperature Corrections}
\vskip 0.1cm
{\Large\bf in the Dilated Chiral Quark Model}
\vskip 0.8in
{\large  Youngman Kim$^{a,c}$, Hyun Kyu Lee$^{a,c}$, Mannque Rho$^{b,c}$}\\
\vskip 0.1in
{\large a) \it Department of Physics, Hanyang University,} \\
{\large \it Seoul 133-791, Korea}\\
{\large b) \it Service de Physique Th\'{e}orique, CEA  Saclay}\\
{\large\it 91191 Gif-sur-Yvette Cedex, France}\\
{\large c) \it Institute for Nuclear Theory, University of Washington}\\
{\large\it Seattle, WA 98195, U.S.A.}
\vskip .4in
\centerline{March, 1995}
\vskip .4in

{\bf ABSTRACT}\\ \vskip 0.1in
\begin{quotation}
\noindent
We calculate the finite-temperature corrections in the dilated chiral
quark model using the effective
potential formalism. Assuming that the dilaton limit is applicable
at some short length scale, we interpret the results to represent
the behavior of hadrons in dense {\it and} hot matter.
We obtain the scaling law,
$\frac{f_{\pi}(T)}{f_{\pi}}  = \frac{m_Q (T)}{m_Q}
 \simeq \frac{m_{\sigma}(T)}{m_{\sigma}}$  while we
argue, using PCAC, that
pion mass does not scale within the temperature range
involved in our Lagrangian. It is found
that the hadron masses and the pion decay constant drop faster with
temperature in the dilated chiral quark model than in the conventional
linear sigma model that does not take into account the QCD scale anomaly.
We attribute the difference in scaling in heat bath to the effect of
baryonic medium on thermal properties of the hadrons. Our finding would imply
that the AGS experiments (dense {\it and} hot matter)
and the RHIC experiments (hot and dilute matter) will ``see" different
hadron properties in the hadronization phase.
\end{quotation}
\end{center}
\end{titlepage}

\newpage

One of the most challenging current topics in hadronic physics
is to understand how physical properties of nuclear matter
change as temperature and density are raised. The most exciting
possibility  is that under extreme conditions of temperature and density,
the matter undergoes one or several phase transitions, such as
meson condensations, chiral symmetry restoration and deconfinement transitions.
Whereas several quark models predict the existence of various
phase transitions, it is the lattice gauge calculations which provided
an unambiguous evidence for the QCD phase transitions.
While the most interesting physics would lie in the quark-gluon or
Wigner-Weyl phase, experiments would necessarily involve hadronization
and hence hadronic
properties in the confined and/or  Goldstone phase. It is therefore
important to understand how the hadrons behave as the system approaches the
critical point from below. For this purpose, the effective Lagrangian
approach has
proven to be quite useful and instructive. Indeed, effective chiral
Lagrangians, constructed to describe the low-energy non-perturbative
sector of QCD, have been used to predict the temperature dependence
of pionic observables\cite{GaL}\cite{hz}. These studies including
lattice calculations were, however,
based on the hadronic vacuum of zero density. The question as to how
the change of vacuum induced by density modifies hadron properties
at finite temperature has not been
addressed in a consistent manner.

A recent development in scaling effect in
effective chiral Lagrangians\cite{Brown} provides a possible means to
address the question: How does dense medium affect the temperature
dependence of hadronic properties?

To answer this question, we use the effective Lagrangian proposed by
Beane and van Kolck\cite{Beane} in the dilaton limit. These authors
incorporate into a nonlinear $\sigma$ model of constituent quarks and
pions the effect of QCD scale anomaly and find that when the ``mended
symmetry" of Weinberg\cite{wein} is taken into account, the effective
Lagrangian ``linearizes" with the scalar field associated with the
trace anomaly becoming a dilaton field $\sigma$. As discussed in \cite{Brown},
this Lagrangian describing the ``mended symmetry" (dilaton) limit (which we
shall refer to as ``dilated chiral quark model") is suitable for physics of
dense hadronic medium with density $\rho$ and would
naturally exhibit the density-dependent scaling proposed in \cite{Rho}.
There is, however, one caveat in this work as well as in work of
ref.\cite{Brown} that should be mentioned. The construction of the chiral
Lagrangian at the dilaton limit leads naturally to a Lorentz-invariant
description of the dynamics. In the presence of matter density, Lorentz
invariance is broken, so the question is how the dilated Lagrangian
should be interpreted when applied to dense matter. We have no clear answer
to this question except that in the dilaton limit, Lorentz-invariant
Lagrangian appears to be a good first approximation. The remaining
question then will be how to apply the dilaton limit to the presence
of dense matter.

In this paper, we study the properties of hadrons at finite
temperature using the Beane-van Kolck Lagrangian. To do this,
we calculate the temperature-dependent
effective potential in the dilated chiral quark model.
We show first that the pion mass in the chiral limit remains zero at
finite temperature as expected
and observe scaling behaviors of the pion decay constant,
$f_\pi$, the $\sigma$ mass and the constituent-quark mass, $m_Q$,
all reflecting the change with temperature of the
vacuum expectation value of the dilaton field $\sigma$. In the Appendix,
we reproduce the results of the one-loop effective potential by calculating
explicitly the finite-temperature mass corrections in coupling-constant
expansion.

The effective chiral quark Lagrangian supplemented with the QCD
conformal anomaly that
Beane and van Kolck\cite{Beane} start with is
\ba
L &=& \bar\psi i(\not\! D + \not\! V ) \psi + g_{A}
\bar \psi\not\! A \gamma_{5}\psi
- \sqrt{\kappa} \frac{m}{f_{\pi}} \bar\psi \psi\chi +\frac{1}{4} \kappa
tr(\partial_{\mu}U \partial^{\mu} U^{\dag})\chi^2\no
 &+& \frac{1}{2} \partial_{\mu}\chi\partial^{\mu}\chi - \frac{1}{2}
tr(G_{\mu\nu}G^{\mu\nu}) -V(\chi)+...
\ea
where $D_\mu = \partial_\mu + igG_\mu$, $V_\mu = \frac{1}{2}i(\xi^\dagger
\partial_\mu\xi + \xi\partial_\mu\xi^\dagger)$ and
$A_\mu = \frac{1}{2}i(\xi^\dagger
\partial_\mu\xi - \xi\partial_\mu\xi^\dagger)$ with $\xi^2 =U=
\exp(\frac{i2\pi_iT_i}{f_\pi})$.
The scale anomaly of QCD appearing at quantum level\cite{Collins}
is contained in the potential $V$ written as
\cite{Josepth}
\be
V(\chi) = -\frac{\kappa m_{\chi}^2}{8f{_\pi}^2 }[\frac{1}{2} \chi^4 -
\chi^4 \ln(\frac{\kappa\chi^2}{f_\pi^2})],
\ee
where $m_{\chi}$ is the mass of the would-be dilaton field, $\chi$.
In the dilated chiral quark model, the dilaton field $\chi$ is
replaced by an effective scalar field $\sigma$ in the linear
basis defined by $\sigma  + i \vec{\tau}\cdot\vec{\pi}= U\chi
\sqrt{\kappa}$.
Near the mended symmetry regime which we assume is realized in dense
medium, the dilaton limit ($m_\sigma \rightarrow 0, \, g_A \rightarrow
1$) is appropriate with both pions and $\sigma$ making up a Goldstone
quartet\cite{Beane}\cite{Georgi}.  In the dilaton limit,
the Lagrangian becomes
\ba
L &=& i\bar Q \not\! \partial Q
+ \frac{1}{2} \partial_\mu \vec{\pi} \partial^\mu \vec{\pi}+
\frac{1}{2}\partial_{\mu}\sigma\partial^{\mu}\sigma
-\frac{m}{f_{\pi}}\bar Q [\sigma -i\gamma_{5}\vec \pi \cdot\vec\tau ]Q \no
 &+& \frac{m_{\sigma}^2}{16 f_{\pi}^2}(\sigma^2 + \vec\pi^2)^2
-\frac{m_{\sigma}^2}{8 f_{\pi}^2}[(\sigma^2 + \vec\pi^2)^2
\ln(\sigma^2 + \vec\pi^2) /f_{\pi}^2].\label{dchmodel}
\ea
Since the minimum of the potential lies on the chiral
circle, $\sigma^2 +\vec{\pi}^2 = f_{\pi}^2$ with $O(4)$ symmetry, we may
choose
\be
<\vec\pi>=\vec\pi_{0}=0 \,\,\,\,
<\sigma>=\sigma_{0}=f_{\pi},\label{vev}
\ee
with the pions remaining as Goldstone bosons.
On this vacuum, the relevant excitations are
\be
m_{\vec\pi}^2 = \frac{\partial^2 V}{\partial\vec\pi^2}\mj = 0 ,\,\,
m_{\sigma}^2=\frac{\partial^2 V}{\partial\sigma^2}\mj =
m_\sigma^2, \,\, m_Q=m.
\ee

We now introduce temperature. This can be done by means of the standard
finite-temperature effective potential\cite{Jackiw}. Define the background
fields by a subscript $c$ and shift the fields as
$\sigma\rightarrow\sigma_{c} + \sigma$ and
$\vec\pi\rightarrow \vec\pi_{c}+ \vec\pi$ in eq.(\ref{dchmodel}). The
Lagrangian takes the form
\ba
L &=& i\bar Q \not\! \partial Q
-\frac{m}{f_{\pi}}\bar Q(\sigma_{c}
-i\gamma_{5}\vec\pi_{c}\cdot\vec\tau )Q \no
&+&\frac{1}{2} \partial_\mu \vec{\pi} \partial^\mu \vec{\pi}+
\frac{1}{2}\partial_{\mu}\sigma\partial^{\mu}\sigma
-\frac{1}{2}\vec\pi M_{\vec\pi}^2\vec\pi -\frac{1}{2}\sigma M_{\sigma}^2\sigma
-\vec\pi M_{\sigma ,\vec\pi}^2\sigma\no
&+& \frac{m_{\sigma}^2}{16 f_{\pi}^2}(\sigma_{c}^2 + \vec\pi_{c}^2)^2
-\frac{m_{\sigma}^2}{8 f_{\pi}^2}(\sigma_{c}^2 + \vec\pi_{c}^2)^2
\ln[(\sigma_{c}^2 + \vec\pi_{c}^2) /f_{\pi}^2]\no
&+& \cdots \label{lc}
\ea
where
\ba
M_{\vec\pi}^2 &=&\frac{\partial^2 V}{\partial\vec\pi^2}\m \no
&=& \frac{m_{\sigma}^2}{2f_{\pi}^2}[(\sigma_{c}^2 +\vec\pi_{c}^2)
\ln((\sigma_{c}^2 +\vec\pi_{c}^2)/f_{\pi}^2)
+2\vec\pi_{c}^2\ln((\sigma_{c}^2 +\vec\pi_{c}^2)/
f_{\pi}^2)+2\vec\pi_{c}^2] \no
M_{\sigma}^2 &=& M_{\vec\pi}^2 (\sigma_{c}\leftrightarrow\vec\pi_{c})\no
M_{\sigma ,\vec\pi}^2&=& \frac{m_{\sigma}^2}{f_{\pi}^2}\sigma_{c}\vec\pi_{c}
(1+\ln((\sigma_{c}^2 +\vec\pi_{c}^2)/f_{\pi}^2).\label{m}
\ea
The ellipsis in eq.(\ref{lc}) stands for the interactions of the
fluctuating fields, $\vec{\pi}$ and $\sigma$, that are not
relevant in calculating the effective potential at one-loop order.
Some of them are explicitly shown in the Appendix.

Let us first calculate the contribution from the fermionic part of
eq.(\ref{lc}). Using the inverse propagator,
\ba
iD^{-1} &=& \not\! k -M_{0} \no
M_{0} &=& \frac{m}{f_{\pi}}(\sigma_{c}
-i\gamma_{5}\vec\pi_{c}\cdot\vec\tau ),
\ea
the contribution to the one-loop finite-temperature effective
potential\cite{Jackiw} can be written as
\ba
V_{Q}^{\beta} &=& i \int \frac{d^4k}{(2\pi)^4} \ln {\mbox
{det}}(\not\! k -M_{0}) \no
            &=& 2i\int \frac{d^4k}{(2\pi)^4} \ln {\mbox {det}}(k^2 -M^2)\no
&=& -4\frac{1}{2\pi^2\beta^4}\int_{0}^{\infty}dx\,x^2
\sum_{i} \,\ln(1+e^{-(x^2+\beta^2M_{i}^2)^{1/2}})\label{vq}
\ea
where $M^2 = \frac{m^2}{f_{\pi}^2}(\sigma_{c}^2+\vec\pi_{c}^2)$
and $\beta= \frac{1}{kT}$.
The first determinant involves Dirac and isospin indices of the
constituent quark, and the second one does only isospin index.
The sum over $i$ goes over the eigenvalues $M_i$ of the matrix $M$.
For small $\beta^2M_{i}^2$ \footnote{Since the masses of the
particles involved in this case are assumed to be  small in the dilaton limit,
the expansion can be used even at quite low temperatures as long as the
temperature is higher than the masses.}
\ba
V_{Q}^{\beta} &\sim& \sum_{i} [\frac{7\pi^2}{180\beta^4}+\frac{1}{12\beta^2}
 \frac{m^2}{f_{\pi}^2}(\sigma_{c}^2 +\vec\pi_{c}^2)] \no
&=& \frac{7\pi^2}{90\beta^4}+\frac{1}{6\beta^2}
 \frac{m^2}{f_{\pi}^2}(\sigma_{c}^2+\vec\pi_{c}^2).\label{vqt}
\ea
Since we are interested in the vacuum structure that depends only on
$\sigma_{c}$ or $\vec\pi_{c}$,  we shall hereafter
retain only those terms in the effective potential that depend upon them
and drop the terms that do not.
Then we get
\ba
V_{Q}^{\beta} &\sim&
\frac{1}{6\beta^2}
 \frac{m^2}{f_{\pi}^2}(\sigma_{c}^2+\vec\pi_{c}^2).\label{vqt1}
\ea

Next we compute the contribution from the meson loops.
Using eq.(\ref{m}), the one-loop finite-temperature contributions
from $\sigma$ and $\vec\pi$ can also be readily calculated\cite{Jackiw}:
\ba
V^{\beta}(\vec\pi_{c}, \sigma_{c})
= &-& \frac{i\hbar}{2} \int\frac{d^4k}{(2\pi)^4}
\ln \,{\mbox {det}}(k^2 -M_{\sigma}^2)\no
&-& \frac{i\hbar}{2}\int\frac{d^4k }{(2\pi)^4 }
\ln \,{\mbox {det}}
(k^2 -M_{\pi}^2 + \frac{(M_{\sigma ,\vec\pi}^2)^2}{k^2-M_{\sigma}^2})\no
=&-&\frac{i\hbar}{2}\int\frac{d^4k}{(2\pi)^4}
\ln \,((k^2)^2 -(M_{\sigma}^2+M_{\vec\pi}^2)k^2+(M_{\sigma ,\vec\pi}^2)^2)\no
= &+& \frac{1}{2\pi^2\beta^4}\int_{0}^{\infty} x^2dx[\ln(1-
e^{-(x^2+\beta^2 R_{1}^2)^{1/2}})\no
&+&\ln(1-e^{-(x^2+\beta^2R_{2}^2)^{1/2}})]
\ea
where $R_{1}^2$ and $R_{2}^2$ are the roots of the quadratic equation in
$k^2$,
$$(k^2)^2 -(M_{\sigma}^2+M_{\vec\pi}^2)k^2+(M_{\sigma ,\vec\pi}^2)^2 =0.$$
For small $\beta^2 R_{i}^2$,
\be
V^{\beta}(\vec\pi_{c},\sigma_{c})\simeq \frac{1}{24\beta^2}(R_{1}^2+R_{2}^2)=
\frac{1}{24\beta^2}(M_{\sigma}^2+M_{\vec\pi}^2).
\ee

Summing up, we have the full one-loop temperature-dependent effective potential
\ba
V_{eff} &=& V_0 + V_{Q}^{\beta}+ V^{\beta}(\vec\pi,\sigma)\no
      &=& -\frac{m_{\sigma}^2}{16 f_{\pi}^2}(\sigma^2 + \vec\pi^2)^2
+\frac{m_{\sigma}^2}{8 f_{\pi}^2}(\sigma^2 + \vec\pi^2)^2
 \ln [\frac{(\sigma^2 + \vec\pi^2)}{f_{\pi}^2}] \no
&+&\frac{1}{6\beta^2}
\frac{m^2}{f_{\pi}^2}(\sigma^2+\vec\pi^2)
+\frac{1}{24\beta^2}(M_{\sigma}^2 +M_{\vec\pi}^2)\label{veff}
\ea
where we have omitted the subscript $c$ in the
$\sigma$ and $\pi$ fields: it should be understood that they are
classical background fields.
Now since chiral symmetry must be intact at finite temperature,
we shall assume that the minimum of the effective
potential, eq.(\ref{veff}), lies on the chiral circle,
$\sigma^2 +\vec{\pi}^2 ={\rm const.}$, as in eq.(\ref{dchmodel}).
Hence in the chiral limit, pions will remain as Goldstone
bosons, with $ <\vec\pi>=0$ and  $<\sigma>=\sigma_0 \neq 0$.
Defining the dimensionless quantities
\ba
x^2=\frac{\vec\pi^2}{f_{\pi}^2}, \,\,
y^2=\frac{\sigma^2}{f_{\pi}^2}, \,\,
r^2=\frac{m^2}{m_{\sigma}^2}, \,\,
t=\frac{1}{\beta f_{\pi}},
\ea
we can rewrite the effective potential as
\ba
V_{eff}&=&\frac{m_{\sigma}^2 f_{\pi}^2}{16}[-(x^2+y^2)^2+2(x^2+y^2)^2
\ln(x^2+y^2)\no
&+& \frac{4}{3}t^2(x^2+y^2)\ln(x^2+y^2)+\frac{2}{3}t^2(x^2+y^2)
+\frac{8}{3}t^2r^2(x^2+y^2)].\label{veff1}
\ea

Now we define the temperature-dependent vacuum $\sigma_0 (T)
$, by extremizing the effective potential as
\ba
\frac{\partial V_{eff}}{\partial\sigma}\mjs=0 , \,\,\,
 \frac{\partial V_{eff}}{\partial\vec{\pi}}\mjs=0,
\ea
and obtain
\be
8y_0^2\ln y_0^2+\frac{8}{3}t^2 \ln y_0^2
+4t^2+\frac{16}{3}t^2r^2 =0,\label{sigma0}
\ee
where $y_0^2={\sigma_0(T)}^2/f_\pi^2$. Below $y$ will correspond
to $y_0$. Equation (\ref{sigma0})
can be solved to get the temperature-dependent vacuum expectation value of
the $\sigma$ field.
One can readily check from eq.(\ref{sigma0}) that in the absence of
temperature, $\sigma_0 = f_{\pi}$ ($y=1$). One can also verify that
the condition (\ref{sigma0}) automatically guarantees that the pion mass
remains zero at all temperatures as required by chiral invariance:
\be
\frac{\partial^2 V_{eff}}{\partial \vec\pi^2}\mjs
\,\,\, = \frac{m_{\sigma}^2}{16}
[8y^2\ln y^2+\frac{8}{3}t^2 \ln y^2 +4t^2+\frac{16}{3}t^2r^2]
\,=\, 0.
\ee
PCAC implies that even when chiral symmetry is broken by the current
quark masses of the u and d quarks, the pion mass will remain small and
unmodified by temperature.

For small $t$, we can approximate
\ba
y=1+\delta, \,\,
\sigma_{0}(T)=f_{\pi}(1+\delta).\label{delta}
\ea
 From eqs.(\ref{sigma0}) and (\ref{delta}), it follows that
\be
\delta = -\frac{1}{4}t^2,\label{deltat}
\ee
for $r^2 \ll 1$.
Now identifying $\sigma_{0}(T)$ with $f_{\pi}(T)$ and using
eq.(\ref{deltat}),  we obtain the scaling of the pion decay constant
\be
\frac{f_{\pi} (T)}{f_{\pi}} =1+\delta=1-\frac{T^2}{4f_{\pi}^2}\label{fpit}
\ee
Next, the $\sigma$ mass is calculated from
the effective potential:
\ba
m_{\sigma}^{2}(T) &=&
\frac{1}{f_{\pi}^2}\frac{\partial^2 V_{eff}}{\partial y^2}
\mjs\no
&=&\frac{m_\sigma^2}{16}[24y^2 \ln y^2 + 16y^2
+\frac{8}{3}t^2 \ln y^2+\frac{28}{3}t^2].
\ea
Using eq.(\ref{delta}) and (\ref{deltat}), we have
\ba
m_{\sigma}^{2}(T) &=& m_\sigma^2[y^2+y^2lny^2+\frac{1}{3}t^2]\no
&\simeq& m_{\sigma}^2(1+\frac{8}{3}\delta) \label{sigmat},
\ea
and hence
\be
\frac{m_\sigma (T)}{m_{\sigma}}
 \simeq 1+\frac{4}{3}\delta =
1-\frac{1}{3}\frac{T^2}{f_\pi^2}
\label{sigmas}
\ee
The temperature-dependent constituent quark mass can be
read off directly from eq.(\ref{dchmodel}):
\ba
m_Q(T) = \frac{m}{f_\pi}\sigma_0(T),\nonumber \\
\ea
or
\ba
 \frac{m_{Q} (T) }{m_Q} &=& 1+\delta= 1-\frac{T^2}{4f_{\pi}^2}\label{mq}.
\ea
We thus find that $m_Q(T)$, $f_{\pi}(T)$ and $m_\sigma (T)$
satisfy approximately the same scaling in temperature as in
density  postulated in \cite{Rho}.

It should be noted that these results are, however, different from
those of $\sigma$ models in matter-free space, namely the results of
chiral perturbation theory at one loop\cite{GaL}, in two
aspects. First, the temperature corrections are much bigger for
two reasons:
the coefficient multiplying $T^2/f_\pi^2$ in eq.(\ref{fpit}) is three times
bigger than in chiral perturbation calculation and furthermore
$f_\pi (\rho)$ in medium is smaller than $f_\pi$ in matter-free space\cite{Rho}
and secondly, the
$\sigma (T)$ scales slightly faster than $f_\pi (T)$ in contrast
to what we expect from the linear $\sigma$ model as we shall show
explicitly below. We can trace the differences to the
presence of the logarithmic-type potential introduced by the
QCD scale anomaly in the dilated chiral model.

In the linear sigma model\cite{Cheng}, the Lagrangian is
\be
L=\frac{1}{2} \partial_\mu \vec{\pi} \partial^\mu \vec{\pi}+
\frac{1}{2}\partial_{\mu}\sigma\partial^{\mu}\sigma
-\frac{1}{8}\frac{m_\sigma^2}{f_\pi^2}[\sigma^2 + \vec\pi^2-f_\pi^2]^2,
\ee
where $\lambda$ and $\mu$ in ref. \cite{Cheng} are replaced by
$f_{\pi}$ and $m_\sigma$,  $f_\pi =(\mu^2 /\lambda)^{1/2}, \,\,
m_\sigma^2 =2\mu^2$.
The finite-temperature effective potential up to one loop is
\ba
V_{eff} = &-&\frac{1}{4}m_\sigma^2(\sigma^2 + \vec\pi^2)
+\frac{1}{8}\frac{m_\sigma^2}{f_\pi^2}(\sigma^2 + \vec\pi^2)^2 \no
&+& \frac{T^2}{24}[\frac{m_\sigma^2}{f_\pi^2}(\sigma^2 + \vec\pi^2)
-\frac{1}{2}m_\sigma^2 ],\label{lveff}
\ea
or using the dimensionless variables,
\ba
V_{eff} = \frac{1}{4}f_\pi^2m_\sigma^2 [-(x^2+y^2)+
\frac{1}{2}(x^2+y^2)^2
+\frac{t^2}{6}(x^2 + y^2 -\frac{1}{2})].
\ea
The vacuum condition, $\frac{\partial V_{tot}^\beta}{\partial y}
= 0$ with $x=0$, leads to
\be
y^2 -1 +\frac{t^2}{6} = 0 .
\label{lsigma0}
\ee
In the low-temperature approximation, eq.(\ref{delta}), we get
\be
\delta (T) \simeq -\frac{T^2}{12f_\pi^2}\label{ldelta}
\ee
and the temperature correction to $f_\pi$:
\be
f_\pi(T) = f_\pi y = f_\pi(1 -\frac{T^2}{12f_\pi^2}).\label{vacu}
\ee
We see that the temperature correction in eq.(\ref{vacu})
is smaller than in eq.(\ref{fpit}). This result is also obtained in chiral
perturbation theory\cite{GaL} (with the nonlinear $\sigma$ model).
The temperature dependence of the $\sigma$ mass in the linear $\sigma$ model
is
\ba
m_\sigma^{2}(T) = -\frac{1}{2}m_\sigma^2 +\frac{3}{2}m_\sigma^2
y^2 +\frac{T^2}{12}\frac{m_\sigma^2}{f_\pi^2}
\ea
which with eq(\ref{lsigma0}) can be reduced to
\be
m_\sigma(T) = m_\sigma \, y = m_\sigma(1 -
\frac{T^2}{12 f_\pi^2})\label{lmsigmat}
\ee
where eq.(\ref{ldelta}) has been used for the last equality. One
can see that the $\sigma$ mass is essentially proportional to
$y$ or $\sigma_0 (T)$ as for $f_\pi(T)$, eq.(\ref{lmsigmat}).
This explains why in the linear sigma model, the ``universal scaling"
\be
\frac{f_\pi(T)}{f_\pi}=\frac{m_\sigma(T)}{m_\sigma}
\ee
holds. In contrast, in the dilated chiral quark model, while
$f_\pi(T)$ is directly proportional to $y$ as it is identified with
$\sigma_0(T)$, $m_\sigma(T)$ is not simply related to
$y$, see eq.(\ref{sigmat}). This is due to the logarithmic term in the
effective potential.

In summary, we have calculated the one-loop effective potential at
finite temperature to determine the temperature dependence of the
light hadron masses (pion, $\sigma$ and constituent quark) and the pion decay
constant, using the dilated chiral quark model in the
dilaton limit. Since in the dilaton limit, the $\sigma$ mass is small,
the low-temperature expansion we are making is justified. We found that
the BR scaling postulated in dense matter also holds approximately in hot
matter in the dilaton limit:
\be
\frac{f_{\pi}(T)}{f_{\pi}}  = \frac{m_Q (T)}{m_Q}
 \simeq \frac{m_{\sigma}(T)}{m_{\sigma}}.
\ee

As stated before, the question remains as to how relevant the dilaton limit on
which the Beane-van Kolck construction relies is
to dense nuclear matter. In the dilaton limit where Weinberg's mended symmetry
applies, the effective Lagrangian preserves Lorentz invariance\footnote{For
instance, in medium, one expects that the pion
decay constant $f_\pi$ (and also the axial coupling constant $g_A$ etc.)
has two components, one for the time component of the
axial current, say, $f_\pi^{(t)}$ and another for the space component,
$f_\pi^{(s)}$. In general, these two are not the same. However in the
dilaton limit, one recovers $f_\pi^{(t)}=f_\pi^{(s)}$.}. But
the presence of the Fermi sea in dense matter
breaks Lorentz invariance, so the valid question is
how good is the assumption
that dense matter can be described by the dilaton limit.
If, however, we assume that the dilated
chiral quark model describes ``vacuum properties" in dense hadronic medium
as argued in \cite{Brown},
the more rapid decrease of $f_\pi$, $m_\sigma$ and $m_Q$ with
temperature compared with free-space $\sigma$ models would imply that
heavy-ion experiments performed in hot and dense medium probe
different hadron properties than in hot but dilute medium. This prediction
could be checked on lattice as well as in heavy-ion experiments.

\subsection*{Acknowledgments}

We are grateful for discussions with U. van Kolck. We would also like
to acknowledge the hospitality of the Institute for Nuclear Theory,
University of Washington where part of this work was done while all three
of us were
participating in the INT95-1 program on ``Chiral dynamics in hadrons and
nuclei."
This work is supported in part by the US Department of Energy,
the KOSEF (Korea) under Grant No. 94-1400-04-01-3
and the Korean Ministry of Education (BSRI-94-2441).
\newpage
\begin{center}
{\bf Appendix }
\end{center}
\setcounter{equation}{0}

In this Appendix, we explicitly calculate the finite-temperature corrections
to the $\sigma$ mass in a coupling-constant expansion up
to order $\frac{m_\sigma^2}{f_\pi^2}$ to check the
results obtained from the one-loop effective potential.
After shifting the fields as $\sigma \rightarrow f_\pi +
\sigma$ and $ \vec{\pi} \rightarrow \vec{\pi}$ in
eq.(\ref{dchmodel}), we get the Lagrangian density:
\ba
L &=& \frac{1}{2} \partial_\mu \vec{\pi} \partial^\mu \vec{\pi}+
\frac{1}{2}\partial_{\mu}\sigma\partial^{\mu}\sigma
-\frac{m}{f_{\pi}} \bar Q [\sigma -i\gamma_{5}\vec \pi \cdot\vec\tau ]Q
+\frac{m_{\sigma}^2}{16 f_{\pi}^2}((f_\pi+\sigma)^2 \no
&+& \vec\pi^2)^2
-\frac{m_{\sigma}^2}{8 f_{\pi}^2}((f_\pi+\sigma)^2 + \vec\pi^2)^2
\ln[1+(\sigma^2 +2f_\pi\sigma +\vec\pi^2) /f_{\pi}^2].\label{dchi2}
\ea
The pion-$\sigma$ interaction vertices can be obtained by expanding the
potential  terms in eq.(\ref{dchi2}),
\ba
L &\simeq&   i\bar Q \not\! \partial Q -m\bar Q Q +
\frac{1}{2} \partial_\mu \vec{\pi} \partial^\mu \vec{\pi}+
\frac{1}{2}\partial_{\mu}\sigma\partial^{\mu}\sigma
-\frac{1}{2}m_{\sigma}\sigma^2\no
&-&\frac{m}{f_{\pi}}\bar Q [\sigma -i\gamma_{5}\vec \pi
\cdot\vec\tau ]Q -\frac{1}{2}\frac{m_{\sigma}^2}{f_\pi}\sigma\vec\pi^2
-\frac{5}{6}\frac{m_{\sigma}^2}{f_\pi}\sigma^3\no
&-&\frac{3}{4}\frac{m_{\sigma}^2}{f_\pi^2}\sigma^2\vec\pi^2
-\frac{1}{8}\frac{m_{\sigma}^2}{f_\pi^2}\vec\pi^4
-\frac{11}{24}\frac{m_\sigma^2}{f_\pi^2}\sigma^4
+ \cdots,
\ea
where only the relevant interactions for finite-temperature corrections
up to order $\lambda =\frac{m_\sigma^2}{f_\pi^2}$ are shown explicitly.

The observation that the $\vec\pi $ field remains massless with
the one-loop finite temperature effective potential can be verified
by checking that there is no temperature-dependent corrections
up to order of $\lambda$.
The relevant diagrams are depicted in Fig.1.
One can show that Fig.1(a)-(e) cancel among themselves and
Fig.1(f) and (g) cancel each other exactly.

The
relevant diagrams for temperature corrections
to the $\sigma$ mass are shown in Fig. 2. They give corrections
to order $\lambda$. A simple calculation gives
\be
\Sigma_{tot}(0) \,=\,\Sigma_{a}(0)\,+\,\Sigma_{b}(0)\,+\,\Sigma_{c}(0)
\,+\,\Sigma_{f}(0)\,=\, -\frac{2}{3}\lambda T^2.
\ee
There are no mass corrections
from Fig. 2 (d) and (e) \cite{Weldon}\cite{Pisa}.
Hence we get
\be
m_\sigma^2 (T) = m_{\sigma}^2 -\frac{2}{3}\lambda T^2
\ee
which is exactly eq.(\ref{sigmat}).

\end{document}